\documentclass[aps,reprint,groupedaddress,amsmath,nofootinbib]{revtex4-1}
\usepackage[colorlinks=true,citecolor=blue,urlcolor=blue,linkcolor=blue]{hyperref}
\usepackage{graphicx}

\begin{document}

%\DeclareGraphicsExtensions{.pdf,.png}

%Title of paper
\title{Collective Yu-Shiba-Rusinov states in magnetic clusters at superconducting surfaces}

\author{Simon K\"{o}rber}
\author{Bj\"{o}rn Trauzettel}
\author{Oleksiy Kashuba}
\email[Email:\,]{okashuba@physik.uni-wuerzburg.de}
\affiliation{Institute for Theoretical Physics and Astrophysics, University of W\"{u}rzburg, Am Hubland, D-97074 W\"{u}rzburg, Germany}

%\date{\today}

\begin{abstract}
We study the properties of collective Yu-Shiba-Rusinov (YSR) states generated by multiple magnetic adatoms (clusters) placed on the surface of a superconductor.
For magnetic clusters with equal distances between their constituents, we demonstrate the formation of effectively spin-unpolarized YSR states with subgap energies independent of the spin configuration of the magnetic impurities.
We solve the problem analytically for arbitrary spin structure and analyze both spin-polarized (dispersive energy levels) and spin-unpolarized (pinned energy levels) solutions.
While the energies of the spin-polarized solutions can be characterized solely by the net magnetic moment of the cluster, the wave functions of the spin-unpolarized solutions effectively decouple from it.
This decoupling makes them stable against thermal fluctuation and detectable in scanning tunneling microscopy experiments.
%\\[-13pt]\begin{center}{\it version~\input{|"git describe --tags"}}\\[-13pt]\end{center}
%\\[-13pt]\begin{center}{\it version~\input{version}}\\[-13pt]\end{center}
\end{abstract}

% insert suggested PACS numbers in braces on next line

% insert suggested keywords - APS authors don't need to do this
%\keywords{}

%\maketitle must follow title, authors, abstract, \pacs, and \keywords
\maketitle

%{\it Introduction}
\section{Introduction}
The progress in understanding the physics of topologically nontrivial systems~\cite{HasanAndKane2010,Moore2010,QiAndZhang2011,Ando2013} has stimulated further research in the field of quantum computation~\cite{Kitaev2003,Nayak2008}.
One reason is that Majorana bound states~\cite{Kitaev2001,Wilczek2009}, which have a topological origin~\cite{Kitaev2009}, can reveal non-Abelian statistics~\cite{Ivanov2001,Kitaev2006,SternAndOppen2004,Stern2010}---a property that can be exploited in topological quantum computing.
Seminal works on the emergence of Majorana bound states are based on $p$-wave superconductivity~\cite{ReadAndGreen2000,Ivanov2001}, but later on it was demonstrated that the same effect can be obtained by the combination of $s$-wave superconductivity, spin-orbit interaction, and modest magnetic fields~\cite{SauAndSarma2010,Alicea2012,Beenakker2013,MourikAndKouwenhoven2012,LutchynAndSarma2010,OregAndOppen2010,Finck2013, Deng2012,DasAndOreg2012,PientkaAndRomito2012}.
Additionally, it has been discovered that a nontrivial topology can also be realized by magnetic adatoms on the surface of $s$-wave superconductors~\cite{PergeAndYazdani2014,LiAndBernevig2014,Dumitrescu2015,Brydon2015,Hui2015,RoentynenAndOjanen2015,FeldmanAndYazdani2016}, where spin-orbit interaction is not necessarily required~\cite{Pientka2013,Nakosai2013,Nadj-Perge2013,ChoyAndBeenakker2011,Martin2012,KlinovajaAndLoss2013,Braunecker2013,Vazifeh2013,Reis2014,Pientka2014,RoentynenAndOjanen2014,PoeyhoenenAndOjanen2014,Kim2014,Christensen2016,Kjaergaard2012,Heimes2014,BrauneckerAndSimon2015,Carroll2017,Schecter2016}.
This is based on the fact that a single magnetic impurity on the surface of an $s$-wave superconductor forms a spin-polarized in-gap state, called the Yu-Shiba-Rusinov (YSR) state~\cite{Yu1965,Shiba1968,Rusinov1969}.
Arranged in a one-dimensional chain, the spins of the impurities interact in this system via Ruderman-Kittel-Kasuya-Yosida (RKKY) interaction~\cite{RudermanAndKittel1954,Kasuya1956,Yosida1957} and align themselves spontaneously in helical order~\cite{KlinovajaAndLoss2013,Braunecker2013,Vazifeh2013,Reis2014,Kim2014,Christensen2016,BrauneckerAndSimon2015,Schecter2016}.
The YSR states in such systems are located close to each other and hybridize, thus forming an in-gap band, and they mimic $p$-wave anomalous correlations, allowing for another possibility of the formation of the Majorana bound states~\cite{Pientka2013,Nakosai2013,Nadj-Perge2013,ChoyAndBeenakker2011,Martin2012,KlinovajaAndLoss2013,Braunecker2013,Vazifeh2013,Reis2014,Pientka2014,RoentynenAndOjanen2014,PoeyhoenenAndOjanen2014,Kim2014,Christensen2016,BrauneckerAndSimon2015,Carroll2017,Schecter2016}.

These discoveries have led to further research on collective YSR states and the physics of magnetic adatoms on the surfaces of superconducting materials~\cite{RubyAndFranke2017,RubyAndFranke2015,RubyAndFranke2016,RubyAndFrankeAndOppen2015,HatterAndFranke2017,Yazdani1997,JiAndXue2008,Heinrich2018,Ptok2017}. It has been demonstrated that the hybridization of YSR states for two impurities leads to novel bound states whose quantum properties can be altered by the distances and local spin orientations between the adatoms~\cite{HoffmanAndLoss2015,Flatte2000,Morr2006,JiAndXue2008,Morr2003,RubyAndFranke2018,Kezilebieke2017,Choi2017,Heinrich2018,Ptok2017}.
We generalize this scenario to a finite set of magnetic impurities (cluster) and derive a theoretical framework for describing the formation of collective YSR states.
If all distances between the magnetic adatoms of the cluster are the same, we find that degenerate, effectively spin-unpolarized YSR states with pinned energy levels arise in the spectrum. These energies are characterized by being robust to the cluster spin configuration (which is experimentally difficult to control).
However, they should be observable by electron spectroscopy because of their robustness.

%The article is organized as follows:
%in Sec.~\ref{sec:Model}., we provide a theoretical model equivalent to the tight-binding BdG Hamiltonian\cite{Pientka2013,Pientka2014}, but written in the basis of impurity spins, which we assume to be classical, playing the role of sites.
%In Sec.~\ref{sec:MagneticCluster}., we consider the magnetic cluster with equal distances between the adatoms, assuming an arbitrary spin configuration.
%Solving this problem analytically, we identify spin-unpolarized solutions whose wave functions effectively decouple from the net magnetic moment of the cluster, ultimately leading to the formation of pinned energy levels in the spectrum of delocalized YSR states.
%Last but not least, we conclude our results in the last section.

%{\it Model}
%\label{sec:Model}

%A microscopic derivation of the hybridization of YSR states and the formation of a band structure inside the superconducting gap was first investigated by Pientka \textit{et al.}~\cite{Pientka2013,Pientka2014}, ultimately leading to a description within the subspace of each YSR state.
\begin{figure}%[htbp]
	\includegraphics[width=\columnwidth]{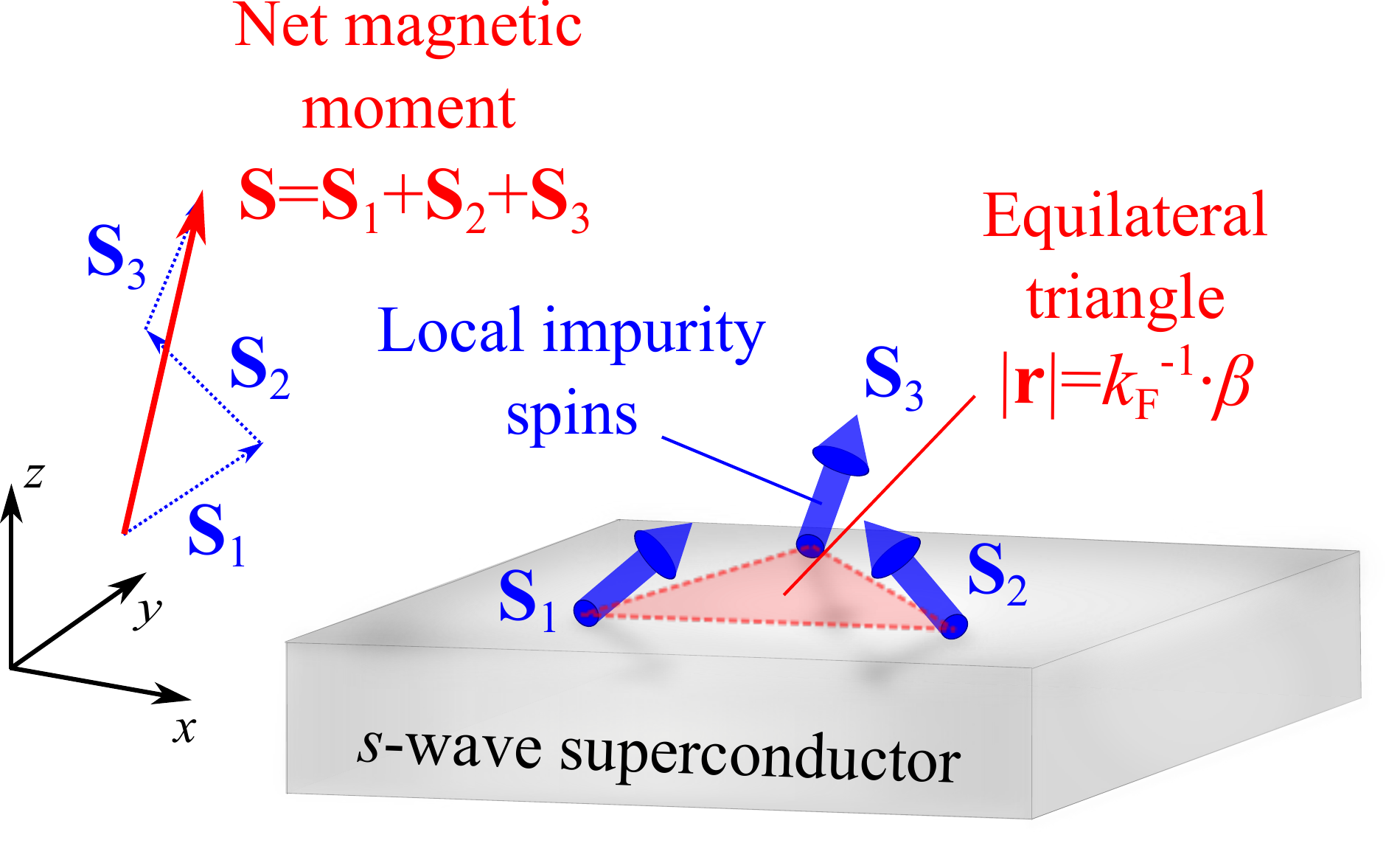}
	\caption{Cluster of $n=3$ magnetic adatoms on the surface of an $s$-wave superconductor.
		The distances $|\mathbf{r}|=k_\text{F}^{-1}\,\beta$ between the impurities are the same, so that the cluster represents an equilateral triangle.
		The magnetic moments of the adatoms are classically described by local impurity spins $\mathbf{S}_i$%
		%with amplitudes $|\mathbf{S}_i|=S_0$
		, while the net magnetic moment $\mathbf{S}=\sum_i\,\mathbf{S}_i$ of the cluster is given by the sum of them.}
	\label{fig:MagneticCluster}
\end{figure}

\section{Model}
For distances $r$ that are much smaller than the coherence length $\xi_0$ of the host superconductor, the indirect exchange couplings between the magnetic adatoms are dominated by RKKY interactions~\cite{RudermanAndKittel1954,Kasuya1956,Yosida1957,BrauneckerAndLoss2009,SimonAndLoss2007,Zener1951,Froehlich1940,Bloembergen1955}, similar to those in a normal metal~\cite{Galitski2002,Aristov1997}.
Then, with the exception of special tunable systems with resonant enhancement of YSR states~\cite{Yao2014}, the static spin texture of the experimentally relevant systems is generally defined by these interactions~\cite{Pientka2013,HoffmanAndLoss2015,PergeAndYazdani2014}. In our work, we take the spin configuration as given without taking into account the processes that determine the orientations of the impurity spins. Then, the adatoms at sites $i$ can be parametrized by fixed spin moments $\mathbf{S}_i$ (see Fig.~\ref{fig:MagneticCluster}) with  absolute values $|\mathbf{S}_i|=S_i$. When the impurities are sufficiently close to each other, YSR states of many adatoms can hybridize~\cite{Pientka2014,Pientka2013,Nakosai2013}, resulting in overlaps that are described by effective transfer integrals between the YSR states. Following the arguments of Refs.~\cite{Nakosai2013,Pientka2013,Pientka2014,ChoyAndBeenakker2011,Martin2012,Nadj-Perge2013}, such system can be represented by an effective Bogoliubov--de Gennes (BdG) lattice model consisting of the magnetic impurities
\begin{equation}
\hat{\text{H}}_\text{eff}=\frac12\sum_{ij}\,\begin{pmatrix}
\hat{\gamma}^\dagger_i & -\hat{\gamma}_i
\end{pmatrix}\,H^{<ij>}_{2\times 2}\begin{pmatrix}
\hat{\gamma}_j\\
-\hat{\gamma}^\dagger_j
\end{pmatrix},
\label{eq:InitialHamiltonian0}
\end{equation}
where
\begin{multline}
	H^{<ij>}_{2\times 2}=
	\begin{pmatrix}
		\epsilon_{i} & 0 \\
		0 & -\epsilon_{i}
	\end{pmatrix}\,\delta_{ij}
	\\
	+\begin{pmatrix}
		-t_{ij}\,\,U^{<ij>}_{\uparrow\uparrow} & \varDelta_{ij}\,\,U^{<ij>}_{\uparrow\downarrow}\\
		\varDelta_{ij}\,\,U^{<ij>}_{\downarrow\uparrow} & t_{ij}\,\,U^{<ij>}_{\downarrow\downarrow}
	\end{pmatrix}(1-\delta_{ij}).
	\label{eq:InitialHamiltonian}
\end{multline}
We have introduced the creation (annihilation) operators $\hat{\gamma}^\dagger_i$ ($\hat{\gamma}_i$) of YSR states at impurity sites $i$, which are spin-polarized along the direction of the impurity spin $\mathbf{S}_i$. The different spin polarizations of YSR states manifest themselves in the spin structure of the transfer amplitudes contained in the $SU(2)$ matrices
\begin{equation}
U^{<ij>}=U^\dagger_iU_j,
\end{equation}
where $U_i$ satisfies the relation $U^\dagger_i(\mathbf{S}_i\cdot\boldsymbol{\sigma})U_i=S_i\sigma_z$ with the vector of Pauli matrices $\boldsymbol{\sigma}=(\sigma_x,\sigma_y,\sigma_z)$. Therefore, the matrices $U^{<ij>}$ carry the information about the impurity spin polarizations at sites $i$ and $j$. The unusual property of the Hamiltonian in Eq.~\eqref{eq:InitialHamiltonian} is that, depending on the mutual orientations of the classical impurity spins, both normal $t_{ij}$ (electron to electron or hole to hole) and anomalous $\varDelta_{ij}$ (electron to hole or hole to electron) hoppings are allowed. Such a Hamiltonian, therefore, depends more on bond properties than on local site states. While Eqs.~\eqref{eq:InitialHamiltonian0} and \eqref{eq:InitialHamiltonian} are universal, the on-site energies $\epsilon_{i}$ and the normal (anomalous) hopping amplitudes $t(\varDelta)_{ij}$ are explicitly dependent on the underlying microscopic model~\cite{Nakosai2013,Ptok2017,Kezilebieke2017} and the low-energy limit of the YSR states~\cite{Pientka2013,Pientka2014}.

A commonly considered BdG Hamiltonian to model YSR states microscopically is given by~\cite{Kondo1964,Pientka2013,Schlottmann1976}
\begin{equation*}
H(\textbf{r})=\xi_{\textbf{p}}\tau_z+\Delta\,\tau_x-\sum_i J\,(\mathbf{S}_i\cdot\boldsymbol{\sigma})\,\delta(\textbf{r}-\textbf{r}_i),
\end{equation*}
where $\xi_\textbf{p}$ is the dispersion relation for the quasiparticles with momentum $\textbf{p}$ in the normal state, and $\Delta$ is the $s$-wave pairing potential of the superconductor.
For simplicity, we assume all spin amplitudes to be the same ($S_{i}=S_{0}$) and neglect the effect of Zeeman splitting\footnote{%
Here we refer to the suppression of the superconducting order parameter in the vicinity of the magnetic impurity.
This suppression is a result of the competition of the singlet pairing and the Zeeman energy of the interaction between each spin of the Cooper pair and the impurity spin.%
} %
from the magnetic impurities~\cite{SalkolaAndSchrieffer1997,HoffmanAndLoss2015,Schlottmann1976,Heinrichs1968,Flatte1997a,Flatte1997b,Flatte2000}.
The Pauli matrices $\tau_n$ $\{n = x, y, z \}$ act in Nambu (electron-hole) space, and $\sigma_n$ in spin space.
The chosen basis of the BdG Hamiltonian is given by the four-component operator $\hat{\Psi}(\textbf{r})=(\hat{\psi}_\uparrow,\hat{\psi}_\downarrow,\hat{\psi}^\dagger_\downarrow,-\hat{\psi}^\dagger_\uparrow)^T$, where $\hat{\psi}_\sigma(\textbf{r})$ are electronic field operators.
The coupling of the superconductor quasiparticles with the spin impurities at positions $\textbf{r}_i$ is controlled by the local exchange coupling strength $J$ and the classical spin moment $\mathbf{S}_i$. We consider the limit of large spin amplitudes $S_0$ and neglect quantum fluctuations of the impurity spins, so that the Kondo effect~\cite{Kondo1964} is suppressed. In this limit, spin-polarized YSR~\cite{Yu1965,Shiba1968,Rusinov1969} states are formed that are quasilocalized at the sites of magnetic impurities.
Each YSR state is characterized by eigenenergies $E_{\pm}=\pm\,\Delta\,(1-\alpha^2)/(1+\alpha^2)$ inside the superconducting gap $\Delta$. We have introduced the local impurity parameter $\alpha=\pi\,\nu_0\,J\,S_0$, where $\nu_0$ is the normal density of states per spin of the host superconductor at the Fermi energy $E_\text{F}$.
The energies $E_{\pm}$ reflect the particle-hole symmetry of the BdG Hamiltonian, resulting in a particle- and hole-like representation of the YSR state at each impurity site. 
For weakly overlapping or deep ($1-\alpha\ll1$) YSR states, the on-site energies are the same and equal to $\epsilon_{0}=\epsilon_{i}\approx\pm E_{\pm}$.
%Assuming vanishing transfer integrals ($t_{ij}=\varDelta_{ij}=0$), the on-site energies of Eq.~\eqref{eq:InitialHamiltonian} are given by the individual YSR energy: $\pm\epsilon_{0}=E_{\pm}$. However, we are interested in delocalized YSR states, which are only present for non-zero transfer amplitudes.
In the deep YSR limit, mathematical expressions simplify~\cite{Pientka2013,Pientka2014}. Then, the normal (anomalous) hopping amplitudes introduced in Eq.~\eqref{eq:InitialHamiltonian} can be written in compact form,
\begin{equation}
t(\varDelta)_{ij}=(-)\Delta \frac{e^{-\frac{\beta_{ij}}{k_\text{F}\xi_0}}}{\beta_{ij}}\sin(\cos)\beta_{ij},
\quad
\epsilon_0\approx(1-\alpha)\Delta.
\end{equation}
Note the dependence on the distances $|\textbf{r}_{ij}|=k_\text{F}^{-1}\,\beta_{ij}$ (where $k_\text{F}$ is the absolute value of the Fermi momentum) between impurities of different sites $i$ and $j$.
%Here, $\epsilon_{i}=\epsilon_0$ is given by the linear approximation of $E_\pm$ about $\alpha=1$, which leads to deep YSR energies $\epsilon_0\approx\Delta\,(1-\alpha)\ll\Delta$ around the center of the superconducting gap.
We emphasize that outside the deep YSR limit, the general structure of Eq.~\eqref{eq:InitialHamiltonian} still holds, with the on-site energy $\epsilon_{0}$ and transfer amplitudes $t(\varDelta)_{ij}$ modified only by global parameters. Therefore, despite the use of the low-energy description~\cite{Pientka2013,Pientka2014}, the following results are not limited to the deep YSR limit. 

The tight-binding BdG Hamiltonian of Eqs.~\eqref{eq:InitialHamiltonian0} and~\eqref{eq:InitialHamiltonian} mixes spin and Nambu (electron-hole) spaces and depends on matrices $U^{<ij>}$ that relate the spin gauges of different sites $i$ and $j$. This dependence results directly from the spin basis of Eq.~\eqref{eq:InitialHamiltonian0}, where the quantization axis of the Nambu operator $\hat{\Psi}(\textbf{r})$ is rotated locally to the orientation $\mathbf{S}_i$ of the corresponding impurity spin~\cite{Kjaergaard2012,ChoyAndBeenakker2011}. Thus, the system is not characterized by natural spin parameters. To provide such a local parameter formulation, we rotate the spin polarizations of the YSR states back to the initial quantization axis of $\hat{\Psi}(\textbf{r})$, which is achieved by a local gauge transformation at each site of the impurities. In particular, we exploit the following idea: we artificially extend the Hilbert space of the YSR states, simplifying the equations by the price of increased dimensionality.
In addition to the YSR states with energies $\epsilon_0$, which we rename as $\hat{\gamma}_j\equiv\hat{\gamma}_{j+}$, we add states of "opposite spin" $\hat{\gamma}_{j-}$, so that there exists a complete and orthonormal set of YSR states at each site~$j$ of the impurities. Then, the extended BdG Hamiltonian based on Eq.~\eqref{eq:InitialHamiltonian} acquires a $4\times4$ matrix structure 
\begin{multline}
H^{<ij>}_{4\times 4}=\left[(\epsilon_0 + \tilde{J}S_{0})\tau_z\otimes\sigma_0-\tilde{J}S_{0}\tau_0\otimes\sigma_z\right] \delta_{ij}
\\
+
\begin{pmatrix}
-t_{ij} & \varDelta_{ij}\\
\varDelta_{ij} & t_{ij}
\end{pmatrix}\otimes U^{<ij>}\,(1-\delta_{ij}),
\label{eq:ExtendedHamiltonian}
\end{multline}
where we have introduced new hopping elements and lifted the energies of the artificially created states $\hat{\gamma}_{j-}$ by $2\tilde{J}S_0$. In the limit $\tilde{J}\to\infty$, the extended Hamiltonian gets projected on the states $\hat{\gamma}_{j+}$ and is reduced to the original one. In Eq.~\eqref{eq:ExtendedHamiltonian}, the basis is given by the discrete four-component Nambu operator $\hat{\Psi}_i=(\hat{\gamma}_{i+},\hat{\gamma}_{i-},\hat{\gamma}^\dagger_{i-},-\hat{\gamma}^\dagger_{i+})^T$, which describes the extended space of the YSR states but still has different spin polarizations at the sites. The benefit of the new Hamiltonian is that Nambu and spin spaces are explicitly decoupled.
Thus, we can perform an $SU(2)$ transformation that rotates the spin space at each site $i$ of the BdG Hamiltonian.
This procedure allows us to obtain an explicit gauge-invariant formulation of our problem,
\begin{multline}
H^{\prime<ij>}_{4\times 4}=(\tau_0\otimes U_i)\,\,H^{<ij>}_{4\times 4}(\tau_0\otimes U_j^\dagger)
\\
=\Big[(\epsilon_0 + \tilde{J}S_{0})\tau_z\otimes\sigma_0-\tilde{J}\tau_0\otimes\mathbf{S}_i\,\boldsymbol{\sigma}\Big]\delta_{ij}
\\
+
\begin{pmatrix}
-t_{ij} & \varDelta_{ij}\\
\varDelta_{ij} & t_{ij}
\end{pmatrix}\otimes\sigma_0\,(1-\delta_{ij}).
\label{eq:GaugeInvariantHamiltonian}
\end{multline}
In this representation, it can be directly seen that the spin degrees of freedom enter into the BdG Hamiltonian only through the on-site terms $\mathbf{S}_i$, while both normal and anomalous hopping terms completely decouple from the spin space of the impurities.
Note, that in contrast to Eq.~\eqref{eq:InitialHamiltonian}, the BdG Hamiltonian of Eq.~\eqref{eq:GaugeInvariantHamiltonian} also allows for the consideration of quantized adatom spins by replacing the classical moment $\mathbf{S}_i$ with the corresponding quantum spin operator.
In this work, however, we do not take the impurity spin dynamics into account.
The classical (static) regime that we consider might be relevant for experiments where the spin amplitudes are mainly given by the $d$ shell with spin $S_0=5/2$ states~\cite{RubyAndFranke2016,RubyAndFrankeAndOppen2015,JiAndXue2008,Heinrich2018,RubyAndFranke2018,Choi2017,RubyAndFranke2015}.
In those cases, the adatom spin configuration is expected to be mostly static.

%{\it Magnetic cluster}
%\label{sec:MagneticCluster}
\section{Magnetic Cluster}

The effects of wave-function hybridization of two magnetic adatoms, such as the formation of bonding and antibonding combinations of YSR states or impurity-induced quantum phase transitions (QPT), have already been investigated in theoretical and experimental studies~\cite{HoffmanAndLoss2015,Flatte2000,Morr2006,JiAndXue2008,Morr2003,RubyAndFranke2018,Kezilebieke2017,Choi2017,Heinrich2018,Ptok2017}. In our work, we focus on phenomena related to quantum interference of YSR states by multiple impurities.
We find that some of the collective wave functions effectively decouple from the net magnetic moment of the adatoms and form pinned energy levels in the spectrum of the magnetic cluster.

In small magnetic clusters on the surface of a superconductor, the adatoms constituting the cluster are on average equidistant and each adatom is coupled to all others.
We can emulate this setup neglecting the random deviations of the couplings $\beta_{ij}$ and take all the distances between the adatoms to be the same, i.e.,\ $\beta_{ij}\equiv\beta_0$ for all pairs $i,j$.
We also assume equal YSR energies $\epsilon_{i}=\epsilon_{0}$ on the equal impurity spins $S_{i}=S_{0}$.
The influence of disorder on these results is discussed in the Appendix~\ref{sm:disorder}.
For the minimal case of three impurities, such a configuration is natural due to van-der-Waals attraction of the magnetic adatoms, which tends to minimize the distances between them.
Such a three-adatom cluster is illustrated in Fig.~\ref{fig:MagneticCluster}.
For the following derivation of the collective YSR states, we consider the more general case of $n\geq3$ magnetic moments of the cluster.\footnote{While the choice $n\geq3$ is mathematically interesting, only the case $n=3$ seems to be experimentally relevant.}

The corresponding BdG equation of the Hamiltonian in Eq.~\eqref{eq:GaugeInvariantHamiltonian} can be solved in the limit of $\tilde{J}\to\infty$ by the amplitudes $\Phi^{\prime\,\text{phy}}_{i}=(a_i\,\chi_i^+,b_i\,\chi_i^-)^T$:
\begin{equation}
\lim_{\tilde{J}\to\infty}\sum_j H^{\prime<ij>}_{4\times 4}\Phi_j^{\prime\,\text{phy}}=E\,\Phi_i^{\prime\,\text{phy}},
%\,\,\,\,\forall\,\,\{i=1,...,n\},
\label{eq:PhysicalLimit}
\end{equation}
where indices $i$ and $j$ run over the impurity index $\{1,\ldots,n\}$ and the spinors $\chi^{\pm}_i$ are the normalized eigenvectors for the Zeeman term at site $i$: $\mathbf{S}_i\,\boldsymbol{\sigma}\,\chi^{\pm}_i=\pm S_0\,\chi^{\pm}_i$. The corresponding coefficients $a(b)_i$ of the physical solutions on the sites $\Phi^{\prime\,\text{phy}}_{i}$ are implicitly determined by the spin-polarized ($\Phi^\prime_0\neq0$) and spin-unpolarized ($\Phi^\prime_0=0$) solutions of the equation of the four-component spinor $\Phi_0^\prime=\sum_i\,\Phi_i^{\prime\,\text{phy}}$ 
[see Appendix~\ref{sm:redmod}, Eq.~\eqref{eq:ReducedModelApp}]: 
%\begin{widetext}
\begin{equation}
\frac{1}{2}\left(A\otimes\sigma_0-B\otimes\mathbf{S}\,\boldsymbol{\sigma} \right)\,\Phi^\prime_0=E\,\,\Phi^\prime_0.
%\\
%A = [2\,\epsilon_0+t_0\,(2-n)]\,\tau_z+\varDelta_0\,n\,\tau_x
%\\
%B= t_0\,\tau_0-i\,\varDelta_0\,\tau_y
\label{eq:ReducedModel}
\end{equation}
In this equation, we define
\begin{align*}
A &= [2\,\epsilon_0+t_0\,(2-n)]\,\tau_z+\varDelta_0\,n\,\tau_x,
\\
B &=[t_0\,\tau_0-i\,\varDelta_0\,\tau_y]/S_0,
%\label{eq:ReducedModel}
\end{align*}
%\end{widetext}
where $t(\varDelta)_{0}=(-)\Delta\,\text{e}^{-\beta_{0}/(k_\text{F}\,\xi_0)}/\beta_{0}\,\sin(\cos)\beta_{0}$ are the normal and anomalous hopping terms and $\mathbf{S}=\sum_i\,\mathbf{S}_i$ is the net magnetic moment of the impurity spins $\mathbf{S}_i$ of the magnetic cluster.

Applying the $SU(2)$ matrices $U^\dagger_i$ to the physical solutions $\Phi^{\prime\,\text{phy}}_{i}$ of Eq.~\eqref{eq:PhysicalLimit}, we effectively rotate them back to the original spin basis.
Removing the components of the artificial states, we then obtain the solutions of the initial Hamiltonian of Eq.~\eqref{eq:InitialHamiltonian}:   
\begin{equation}
\sum_j\,H^{<ij>}_{2\times 2}\,\Phi_j^{\text{ini}}=E\,\,\Phi_i^{\text{ini}},
\qquad
\Phi^\text{ini}_i=\begin{pmatrix}
a_i\\
b_i
\end{pmatrix}.
\end{equation}

\textit{Spin-polarized solutions.} For spinors $\Phi^\prime_0\neq0$, the amplitudes $\Phi^{\prime\,\text{phy}}_{i}$ can be explicitly expressed through $\Phi_0^\prime$ [see Appendix~\ref{sm:redmod}, Eq.~\eqref{eq:SpinorsImpuritySiteupdown}] 
yielding the coefficients $a(b)_i$.
The reduced Eq.~\eqref{eq:ReducedModel} contains only the net magnetic moment $\mathbf{S}$ of the cluster. Consequently, if we find solutions $\Phi_0^\prime\neq0$ of Eq.~\eqref{eq:ReducedModel}, the spin components of these spinors are given by the spin-up and spin-down solutions of $\mathbf{S}\,\boldsymbol{\sigma}$, which is why we call them spin-polarized states.
Considering additionally the Nambu space, this leads to four spin-polarized solutions $\Phi_0^\prime$, which represent dispersive energy levels that depend solely on $(\mathbf{S}\cdot\boldsymbol{\sigma})^2=|\mathbf{S}|^2$.

\textit{Spin-unpolarized solutions.} Since Eq.~\eqref{eq:InitialHamiltonian} has $2\,n$ solutions, $2\,n-4$ are still missing.
These are the solutions corresponding to the spin-unpolarized case $\Phi^\prime_0=0$.
In this case, $\Phi_{i,+}^{\text{tri}}=(a^+_i,0)^T$ and $\Phi_{i,-}^{\text{tri}}=(0,b^-_i)^T$ are the solutions of the initial Hamiltonian, where the corresponding energy levels are given by $E_\pm^{\text{tri}}=\pm(\epsilon_0+t_0)$ [see Appendix~\ref{sm:redmod}].
Here, the coefficients $a_i^+$ and $b_i^-$ satisfy the conditions
\begin{equation}
\sum_i\,a_i^+\,\chi_i^+=0\,,\,\,\sum_i\,b_i^-\,\chi_i^-=0.
\label{eq:SpinUnpolarized}
\end{equation}
Each of these conditions contain $n-2$ linearly independent and normalized sets of nontrivial coefficients $a_i^+$ and $b_i^-$, which means that each pinned energy level $E_\pm^{\text{tri}}$ is $n-2$ times degenerate.

\begin{figure}%[htbp]
	\includegraphics[width=\columnwidth]{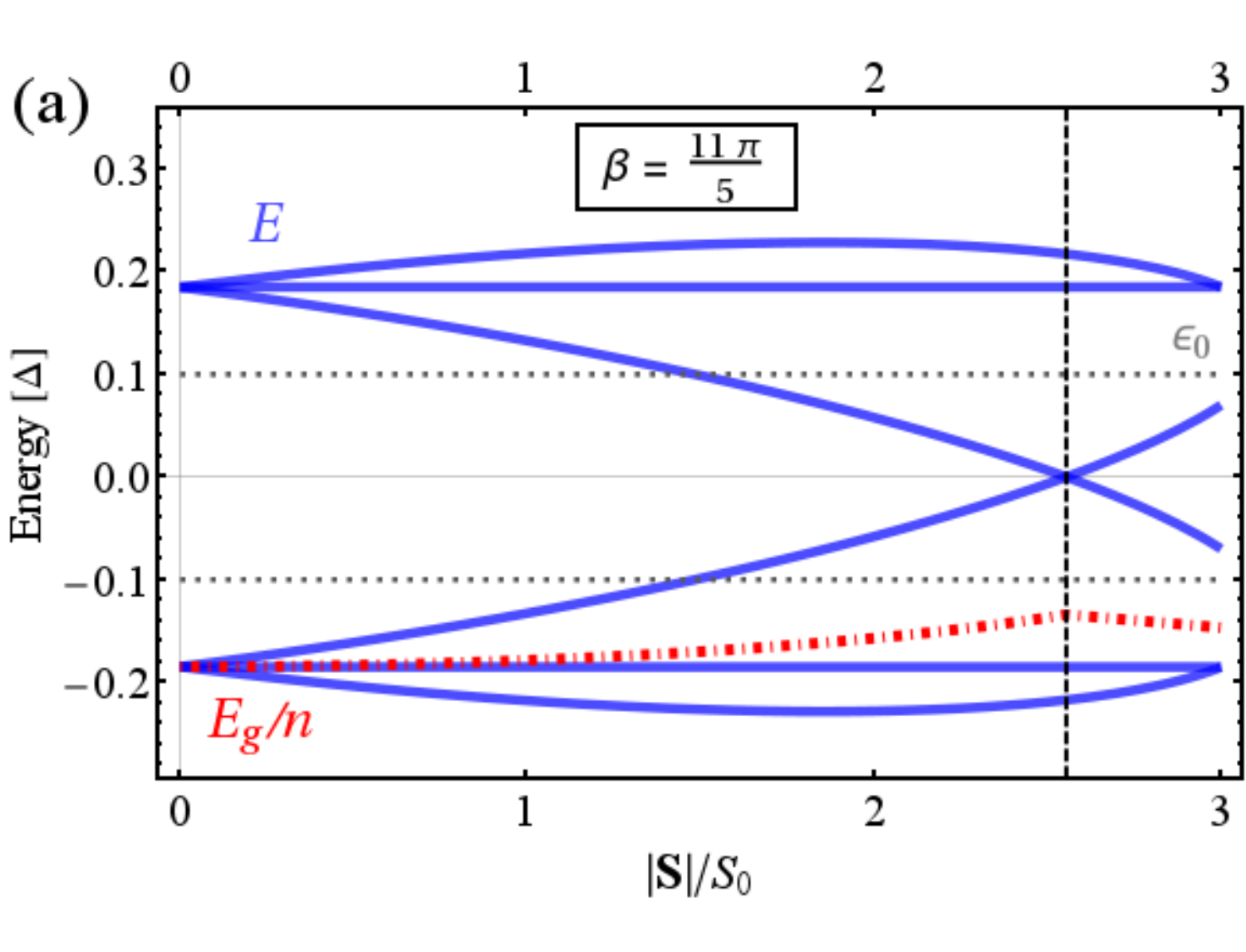}\\
	\includegraphics[width=\columnwidth]{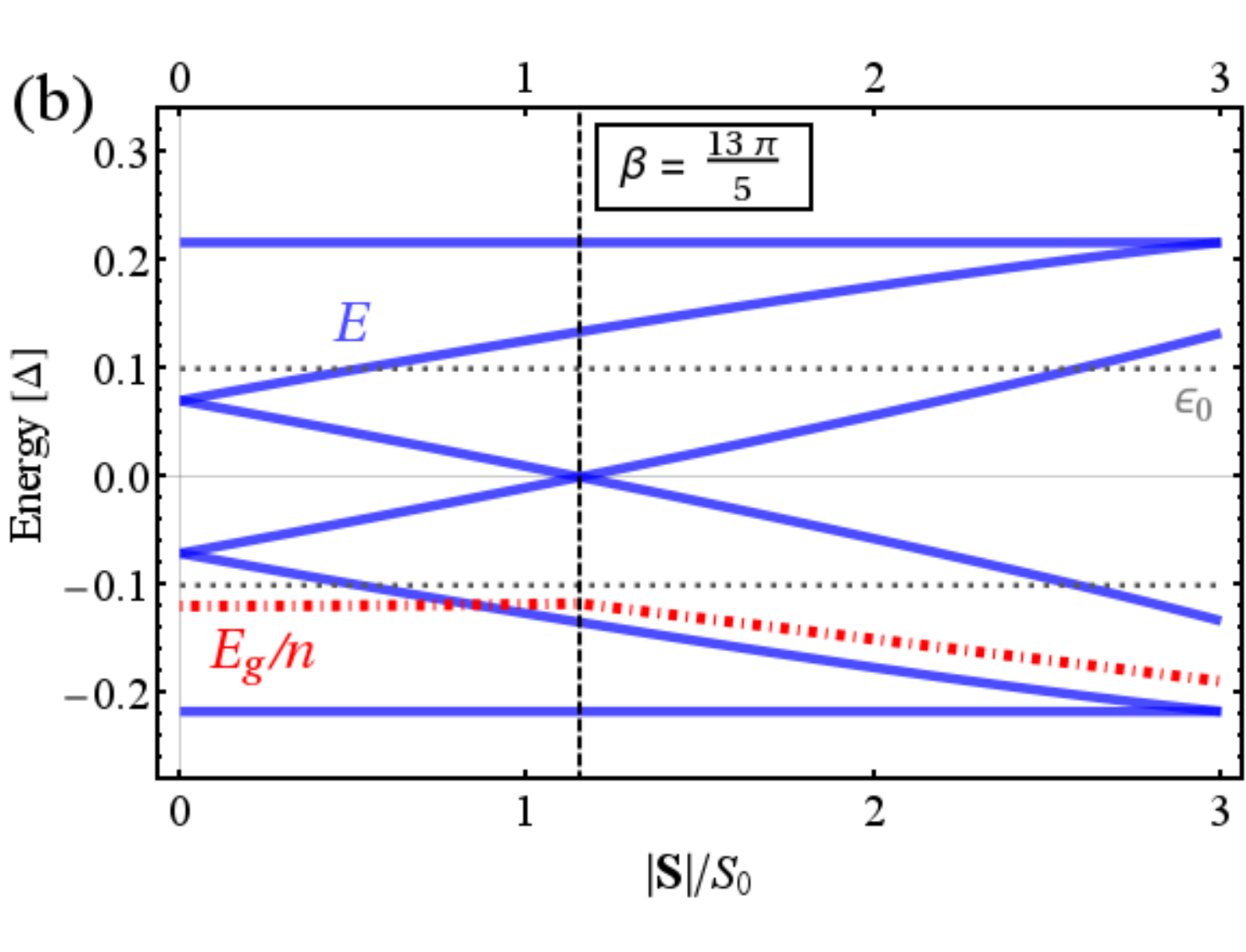}
	\caption{Energy levels $E$ (solid, blue) of the magnetic cluster of $n=3$ impurities for $\alpha=0.9$ and distances $|\textbf{r}|=k_\text{F}^{-1}\,\beta\ll\xi_0$ between the adatoms.
		The ground-state energy $E_\text{g}$ (dashed red line) is represented in dependence on the net magnetic moment $|\mathbf{S}|$ of the cluster.
		In addition to four dispersive energy levels, two $n-2$ degenerate and pinned energy levels emerge in the spectrum for both parameters (a) $\beta=\frac{11\,\pi}{5}$ and (b) $\beta=\frac{13\,\pi}{5}$.
		The vertical dotted lines denote the moment $S_\text{c}$ where the quantum phase transition takes place.
		The horizontal dotted lines show the position of the deep Yu-Shiba-Rusinov energy $\epsilon_{0}$ at an isolated impurity.}
	\label{fig:EnergyPlots}
\end{figure}

We illustrate the typical spectrum of the cluster using the three-impurity setup shown in Fig.~\ref{fig:EnergyPlots}.
We consider deep YSR states with impurity parameter $\alpha=0.9$ and demonstrate the energy level dependence on the only relevant magnetic configuration parameter---the net magnetic moment $|\mathbf{S}|$ of the cluster.
%In Fig.~\ref{fig:EnergyPlots}, the delocalized YSR states (blue, solid lines) for $n=3$ impurities and a Kondo parameter $\alpha=0.9$ are represented in dependence on the net magnetic moment $|\mathbf{S}|=S$ of the cluster.
Additionally, we draw the dependence of the net ground-state energy $E_\text{g}=\sum_{E_i\leq0}\,E_i$ (red, dashed line), which may indicate the preferred spin configuration in the case when the YSR exchange dominates over the RKKY interaction.
This can happen at special values of $\beta$~\cite{Yao2014}. % within the description of the YSR states.
The pinned energy levels do not depend on the magnetic cluster configuration for any distance $|\textbf{r}|=k_\text{F}^{-1}\,\beta$.
This feature follows naturally from the fact that the associated spin-unpolarized solution effectively decouples each site from the others, as we have shown in Eqs.~\eqref{eq:ReducedModel} and~\eqref{eq:SpinUnpolarized}.
This decoupling happens due to the exact compensation of normal and anomalous hopping between the sites visible through the site gauge rotation in Nambu space [see Appendix~\ref{sm:redmod}, Eq.~\eqref{eq:AppBdGEquation}]
This applies as long as the separations between impurities are identical, while random deviations influence the polarization of the solutions and the pinned energy levels [see Appendix~\ref{sm:disorder}].
In the particular case of $n=3$, the two pinned levels are non-degenerate.
Additionally, the four spin-polarized solutions corresponding to dispersive energy levels of the hybridized YSR states are illustrated.
For the particular parameters in Fig.~\ref{fig:EnergyPlots}, we can observe the dispersive levels crossing at zero energy.
This point corresponds to a QPT~\cite{Sakurai1970,SalkolaAndSchrieffer1997}, where the fermionic parity of the groundstate changes~\cite{Balatsky2006,Morr2006,HoffmanAndLoss2015,Morr2003}.
In the YSR exchange-dominated case, the net ground energy $E_\text{g}$ determines the cluster configuration.
Depending on the distance between the adatoms of the magnetic cluster, this may either result in ferromagnetic (at $\beta=\frac{13\,\pi}{5}$, and $|\mathbf{S}|=3\,S_0$) or hedgehog-like (at $\beta=\frac{11\,\pi}{5}$ and $|\mathbf{S}|=0$) configurations of the impurity spins.

%{\it Conclusions}
\section{Summary}

We have introduced a theoretical model that describes the coupling between Yu-Shiba-Rusinov states and depends only on the local spins $\mathbf{S}_i$ of the adatoms.
By analyzing a magnetic cluster of $n$ impurities, we have reduced the model to a simplified Bogoliubov--de Gennes equation, where we have identified both spin-polarized and spin-unpolarized solutions.
These solutions lead to (a) four dispersive energy levels and (b) two $(n-2)$-degenerate pinned energy levels, which are robust against the net magnetic moment.
The dispersive energies can be characterized solely by the net moment $|\mathbf{S}|$ of the magnetic cluster, and they can experience a quantum phase transition associated with the fermionic parity of the ground state.

\begin{acknowledgments}
Financial support by the DFG (SPP1666 and SFB1170 "ToCoTronics") and the ENB graduate school on Topological Insulators is gratefully acknowledged.
We thank S.~Nakosai and K.~Franke for stimulating discussions.
\end{acknowledgments}

\appendix

\section{Disorder effects}
\label{sm:disorder}

\begin{figure}%[htbp]
	\includegraphics[width=0.9\columnwidth]{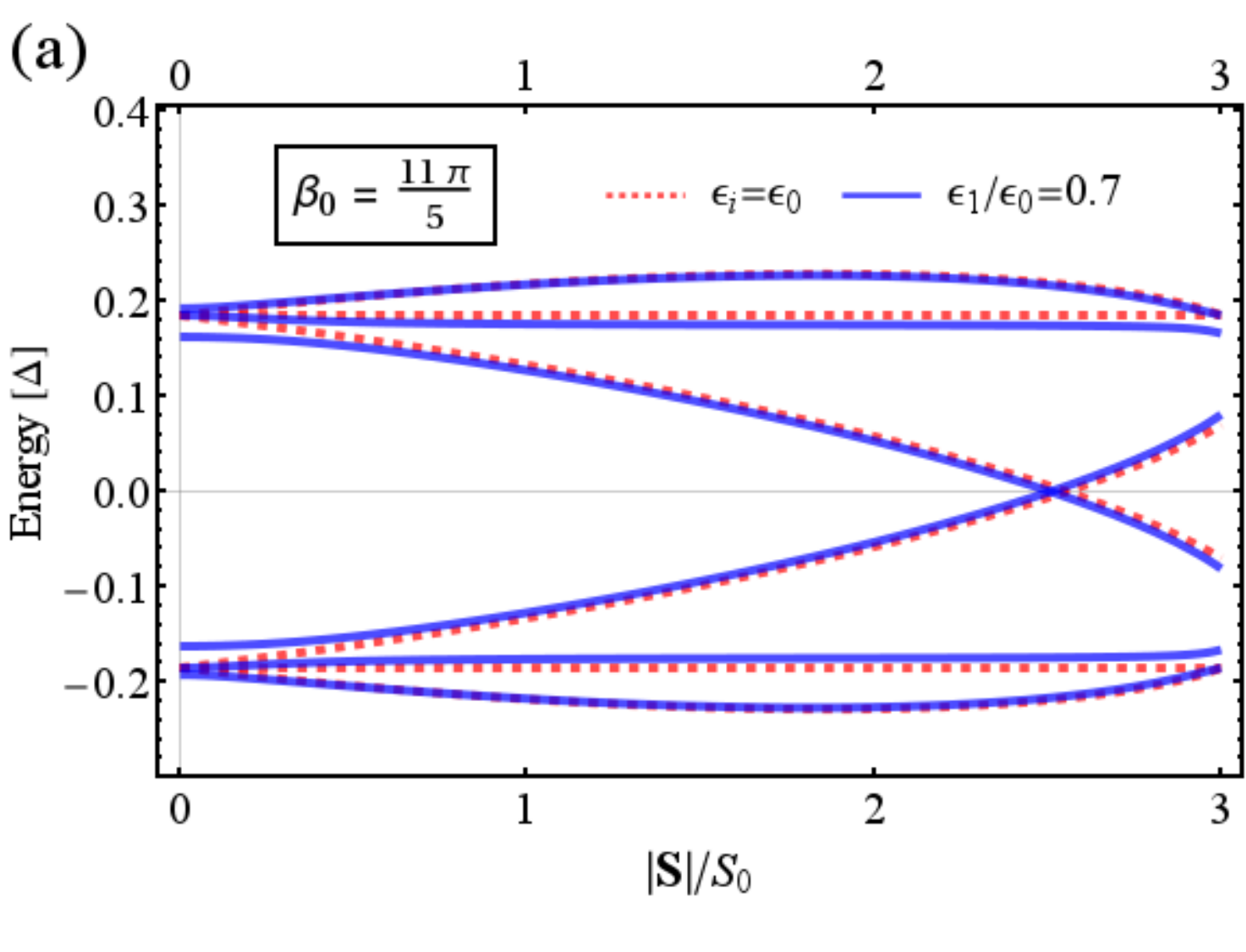}\\
	\includegraphics[width=0.9\columnwidth]{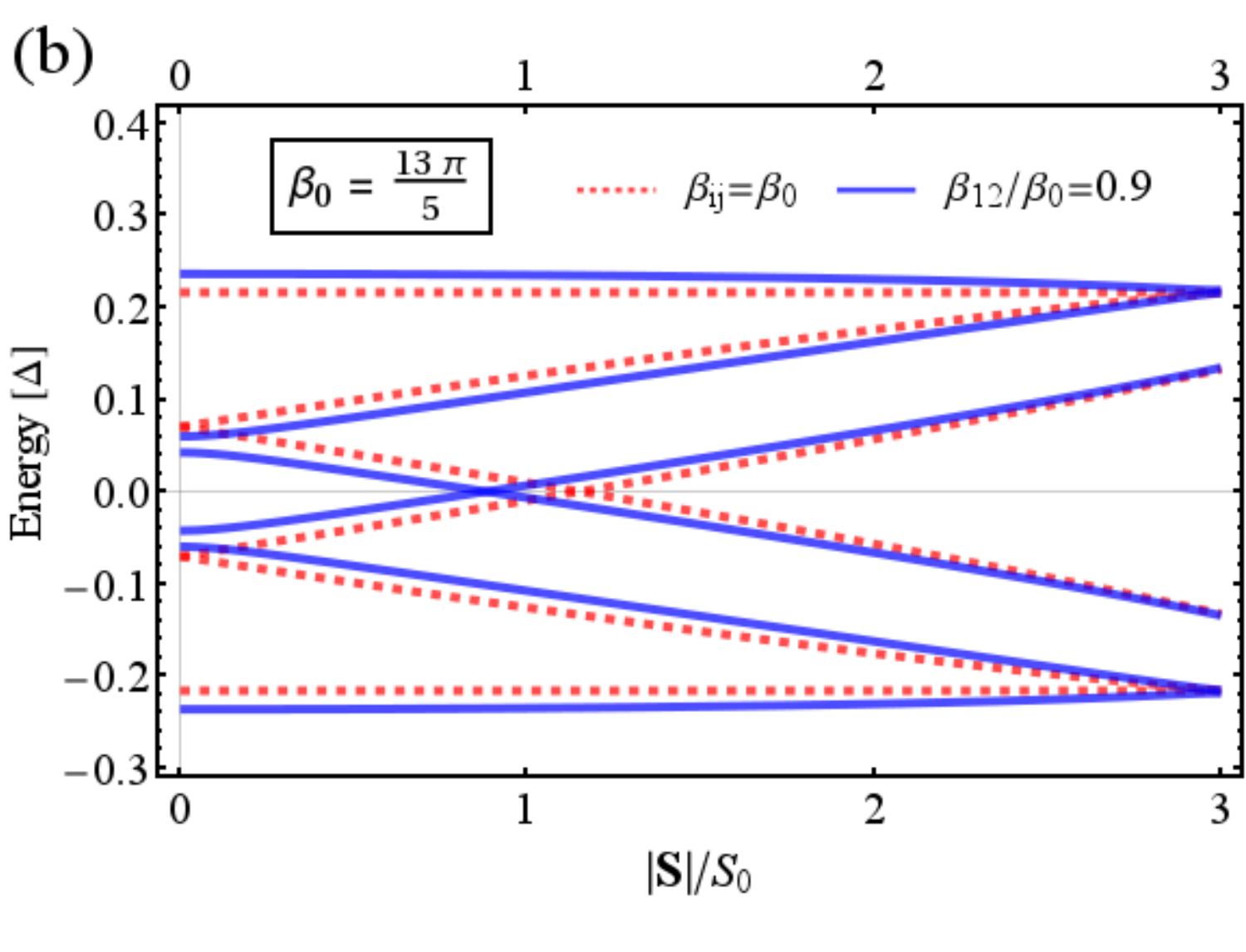}
	\caption{Energy levels of the magnetic cluster of three impurities.
(a) Comparison of the energy levels of the symmetric cluster from Fig.~\ref{fig:EnergyPlots}a) of the main text [red dashed lines] with the same cluster with one on-site energy changed to $\epsilon_{1}=0.7\epsilon_{0}$ [blue lines].
(b) Comparison of the energy levels of the symmetric cluster from Fig.~\ref{fig:EnergyPlots}b) of the main text [red dashed lines] with the same cluster with one bond length changed to $\beta_{12}=0.9\beta_{0}$ [blue lines].}
	\label{fig:DisorderPlots}
\end{figure}

The spin-unpolarized solutions we referred to in the main text [see details in Appendix~\ref{sm:redmod}] have been derived under the assumption of equal on-site energies $\epsilon_{i}=\epsilon_{0}$ and equidistant placement of the impurities $\beta_{ij}=\beta_{0}$.
This choice implies equal tunneling amplitudes $t_{ij}$ and $\varDelta_{ij}$.
We have studied numerically the effect of disorder on the three-spin cluster varying either one of the on-site energies (potential disorder) or one of the distances (positional disorder).
The result is presented in Fig.~\ref{fig:DisorderPlots}.

\newpage

\section{Derivation of the reduced model}
\label{sm:redmod}

The BdG equation of the gauge-invariant Hamiltonian of Eq.~\eqref{eq:GaugeInvariantHamiltonian} is generally given by the following expression:
\begin{equation}
\sum_j\,H^{\prime<ij>}_{4\times 4}\,\Phi_j^\prime=E\,\Phi_i^\prime\qquad\forall\,\,\{i=1,...,n\},
\end{equation}
where $n$ is the number of magnetic impurities.
Here, each spinor $\Phi_i^\prime$ represents the solution of the system at the individual site $i$ and is expressed within a discrete four-component Nambu space.
Parametrizing the spinors as $\Phi^\prime_i=\Phi^{\prime\,+}_i+\Phi^{\prime\,-}_i$, where $\Phi^{\prime\,\pm}_i=\phi^{\prime\,\pm}_i\otimes\chi_i^\pm$, the BdG equation for $n\geq2$ impurities arranged in the magnetic cluster ($\beta_{ij}\equiv\beta_0=\text{const.}$) yields
\begin{widetext}
\begin{equation}
\begin{pmatrix}
\epsilon_0+t_0 & -\varDelta_0\\
-\varDelta_0 & -2\tilde{J}S_{0}-\epsilon_0-t_0
\end{pmatrix}\,\Phi^{\prime\,+}_i+\begin{pmatrix}
2\tilde{J}S_{0}+\epsilon_0+t_0 & -\varDelta_0\\
-\varDelta_0 & -\epsilon_0-t_0
\end{pmatrix}\,\Phi^{\prime\,-}_i
+
\begin{pmatrix}
-t_0 & \varDelta_0\\
\varDelta_0 & t_0
\end{pmatrix}\,\Phi^\prime_0=E\,(\Phi^{\prime\,+}_i+\Phi^{\prime\,-}_i).
\label{eq:AppBdGEquation}
\end{equation}
\end{widetext}
Here, we have introduced the four-component spinor
\begin{equation}
\Phi_0^\prime=\sum_i\,\Phi_i^\prime,
\label{eq:AppDefPhi}
\end{equation}
which will later play a key role in interpreting the solutions.

For the spin-unpolarized case $\Phi_0^{\prime}=0$, Eq.~\eqref{eq:AppBdGEquation} is directly solved in the limit $\tilde{J}\to\infty$ by a separation ansatz: $\Phi^{\prime}_{i,+}=
(a_i^+\,\chi_i^{+},
0)^T$ and $\Phi^{\prime}_{i,-}=(
0,
b_i^-\,\chi_i^{-}
)^T$, where the corresponding energies yield $E^\text{tri}_\pm=\pm(\epsilon_0+t_0)$.
Thus, the coefficients $a_i^+(b_i^-)$ satisfy the conditions $\sum_i\,a_i^+(b_i^-)\,\chi_i^{+}(\chi_i^{-})=0$, leading to $n-2$ linear independent and normalized sets of $a_i^+(b_i^-)$.
However, for spin-polarized solutions $\Phi^\prime_0\neq0$, we multiply Eq.~\eqref{eq:AppBdGEquation} by $\tau_0\otimes P_i^\pm$, where $P_i^\pm=\chi_i^\pm(\chi_i^\pm)^\dagger$ is the projector of the corresponding spin-up and spin-down solutions $\chi_i^\pm$ of the local impurity spins $\mathbf{S}_i$.
Taking the limit $\tilde{J}\to\infty$, we find an explicit expression for the amplitudes on the sites $\Phi^{\prime\,\pm}_i$ through $\Phi^\prime_0$:
\begin{subequations}
\begin{align}
%\text{I}.)\,\,\,
\Phi^{\prime\,+}_{i}&=\frac{1}{-E+t_0+\epsilon_0}\,\begin{pmatrix}
t_0 & -\varDelta_0\\
0 & 0
\end{pmatrix}\otimes P_i^+\,\,\Phi^\prime_{0},
\label{eq:SpinorsImpuritySiteup}
%\qquad\text{and}\qquad
%\end{align}
%and
%\begin{align}
%\text{II}.)\,\,\,
\\
\Phi^{\prime\,-}_{i}&=\frac{1}{E+t_0+\epsilon_0}\,\begin{pmatrix}
0 & 0\\
\varDelta_0 & t_0
\end{pmatrix}\otimes P_i^-\,\,\Phi^\prime_{0}.
\label{eq:SpinorsImpuritySitedown}
\end{align}
\label{eq:SpinorsImpuritySiteupdown}
\end{subequations}
Summing over the sites $i$ and spins, such as in Eq.~\eqref{eq:AppDefPhi}, we obtain an equation for the spinor $\Phi^\prime_0$:
\begin{align}
\Phi^\prime_0=\Bigg[\frac{1}{-E+t_0+\epsilon_0}\,\begin{pmatrix}
t_0 & -\varDelta_0\\
0 & 0
\end{pmatrix}&\otimes P^+
%\notag\\
%&\hspace*{1.96cm}
\nonumber\\
+\frac{1}{E+t_0+\epsilon_0}\,\begin{pmatrix}
0 & 0 \\
\varDelta_0 & t_0
\end{pmatrix}&\otimes P^-\Bigg]\,\,\Phi^\prime_0,
\end{align}
where the matrices $P^\pm=\sum_i\,P_i^\pm=\frac{1}{2\,S_0}\,\Big(S_0\,n\pm\mathbf{S}\,\boldsymbol{\sigma}\Big)$ are written in terms of the net magnetic moment $\mathbf{S}=\sum_i\,\mathbf{S}_i$.
Finally, this equation can be rewritten into compact form
\begin{equation}
\begin{pmatrix}
\epsilon_0+t_0\,(1-P^+) & \varDelta_0\,P^+\\
\varDelta_0\,P^- & -\epsilon_0-t_0\,(1-P^-)
\end{pmatrix}\,\Phi_0^\prime=E\,\Phi^\prime_0.
\label{eq:ReducedModelApp}
\end{equation}
By finding a spin-polarized solution $\Phi^\prime_0\neq0$ of this equation, the corresponding spinors $\Phi^{\prime\,\pm}_i$ at the impurity sites can now be calculated immediately by Eqs.~\eqref{eq:SpinorsImpuritySiteup} and~\eqref{eq:SpinorsImpuritySitedown}, 
which leads to the parametrization
\begin{equation}
\Phi^{\prime\,\text{phy}}_{i}=
\begin{pmatrix}
a_i\,\chi_i^+\\
b_i\,\chi_i^-
\end{pmatrix}
\end{equation}  
of the physical solutions.
Thus, the coefficients $a(b)_i$ can be explicitly determined by solving Eq.~\eqref{eq:ReducedModelApp}.
Furthermore, we refer to the solutions as physical, as the limit $\tilde{J}\to\infty$ was made before the identification of the spinors $\Phi^{\prime\,\text{phy}}_{i}$.

%\newpage

% Create the reference section using BibTeX:
\bibliographystyle{apsrev4-1}
\bibliography{ShibaBands}

\end{document}